\def\mearth{{\rm\,M_\oplus}}
\def\rearth{{\rm\,R_\oplus}}
\def\msun{{\rm\,M_\odot}}

\def\gsim{~\rlap{$>$}{\lower 1.0ex\hbox{$\sim$}}}
\def\lsim{~\rlap{$<$}{\lower 1.0ex\hbox{$\sim$}}}
\def\etal{{\it et al.\thinspace}}
\def\wpm2{W m$^{-2}$}
\def\etal{{\it et al.\thinspace}}
\def\eg{{\it e.g.\ }}

\def\ie{{\it i.e.\ }}

\documentclass[12pt, preprint]{aastex}
\usepackage{graphicx}
\usepackage{natbib,color,lscape}

\title{CoRoT-7 b: Super-Earth or Super-Io?}
\author{Rory Barnes\altaffilmark{1,2}, Sean
N. Raymond\altaffilmark{2,3}, Richard Greenberg\altaffilmark{4}, Brian
Jackson\altaffilmark{5,6}, Nathan A. Kaib\altaffilmark{1}}

\altaffiltext{1}{Department of Astronomy, University of Washington, Seattle,
WA, 98195-1580}
\altaffiltext{2}{Virtual Planetary Laboratory}
\altaffiltext{3}{Center for Astrophysics and Space Astronomy, University of
 Colorado, UCB 389, Boulder, CO 80309-0389}
\altaffiltext{4}{Lunar and Planetary Laboratory, University of Arizona, Tucson, AZ 85721}
\altaffiltext{5}{Planetary Systems Laboratory, Goddard Space Flight
Center, Code 693, Greenbelt, MD 20771}
\altaffiltext{6}{NASA Postdoctoral Program Fellow}

\begin{document}
\begin{abstract}
CoRoT-7 b, a planet about 70\% larger than the Earth orbiting a
Sun-like star, is the first-discovered rocky exoplanet, and hence has
been dubbed a ``super-Earth''. Some initial studies suggested that since
the planet is so close to its host star, it receives enough insolation
to partially melt its surface. However, these past studies failed to
take into consideration the role that tides may play in this system. Even
if the planet's eccentricity has always been zero, we show that tidal
decay of semi-major axis could have been large enough that the planet
formed on a wider orbit which received less insolation. Moreover,
CoRoT-7 b could be tidally heated at a rate that dominates its
geophysics and drives extreme volcanism.  In this case, CoRoT-7 b is a
``super-Io'' that, like Jupiter's volcanic moon, is dominated by
volcanism and rapid resurfacing.  Such heating could occur with an
eccentricity of just $10^{-5}$. This small value could be driven by
CoRoT-7 c if its own eccentricity is larger than $\sim 10^{-4}$. CoRoT-7 b
may be the first of a class of planetary super-Ios likely to be
revealed by the
\textit{CoRoT} and \textit{Kepler} spacecraft. 
\end{abstract}

\keywords{planets and satellites: individual (CoRoT-7 b) -- celestial mechanics}

\section{Introduction}

The discovery of CoRoT-7 b (L\'eger, Rouan, Schneider \etal 2009;
hereafter LRS09) heralds a new era in
the study of exoplanets. The detection of rocky exoplanets has been a prime objective of
exoplanet surveys as they are the most likely environments to support
life. With a radius $R_p \sim 1.7 \rearth$
(LRS09), and a mass of 4.8
$\mearth$ (Queloz \etal 2009), this planet is likely rocky (Valencia
\etal 2007; Fortney \etal 2007; Sotin \etal 2007; Seager \etal
2007). CoRoT-7 b is clearly not habitable, being only 0.017 AU from
its host star, but understanding its origin and properties can inform
future interpretations of rocky exoplanets with the potential to be
habitable.

Considerable research has already explored the plausible properties of
close-in ``super-Earths'', rocky planets with masses $\lsim 10
\mearth$ and semi-major axes $a \lsim 0.1$ AU (see Gaidos \etal 2007 for a review). Such planets probably formed at larger
distances and migrated in, although there exist several other
formation mechanisms (Raymond \etal 2008). At very close distances ($\lsim
0.03$ AU), planets orbiting $\sim 1 \msun$ stars may experience enough
insolation to partially melt the surface, producing a silicate
vapor atmosphere (Schaefer \& Fegley 2009). CoRoT-7 b may be such a
world with a surface temperature of order 2000 K (LRS09,
Valencia \etal 2009).

These previous studies assumed the planet remains at a constant distance
from the star.  However, CoRoT-7 b is so close to its star that its
life expectancy is very short as tides cause the orbit to decay
quickly, assuming a conventional value for the stellar tidal parameter
$Q'_*$ (Jackson \etal 2009).  On that basis, LRS09 inferred that the
value of $Q'_*$ is unexpectedly large.  Alternatively, CoRoT-7 b could
be on the verge of spiralling down to
the star, out of a large population of planets (in many systems), many
of which have already been destroyed, as discussed by Jackson
\etal (2009). It therefore remains uncertain whether the radiative heating
has been intense enough, long enough to melt the surface. Here we
assess the possible orbital history and its implications.

Tides may also contribute to, and even dominate, the heating of the
planet.  If the planet's orbit is at all eccentric (even if the
eccentricity $e_b$ is $< 0.03$ so as to be extremely difficult to
measure via radial velocity data [Butler \etal 2006]), the planet's
figure is flexed by the time-varying gravitational potential of
the star.  These tides on the planet transform orbital energy into
internal heat, affecting its geophysics as well as contributing to
orbital decay. If the orbit of CoRoT-7 b is sufficiently eccentric,
then the planet could be substantially affected by tidal heating.
Tidal heat alone could be adequate to make it a ``super-Io'', with a
surface heat flux far exceeding that of Jupiter's extremely volcanic
moon (Barnes \etal 2009a). This paper identifies the orbital
circumstances, histories, and physical properties that lead to Io-like
volcanism (a super-Io) and those that do not (a super-Earth).

If CoRoT-7 b orbited in isolation, then tides would quickly damp
any primordial eccentricity (LRS09), and we might expect minimal
tidal heating today. However, even small eccentricities (those
typically considered negligible) could yield considerable tidal
heating and have profound consequences for
the geophysical state of this planet. We show here that if the planet
is similar to the rocky bodies in our Solar System and has an
eccentricity even as small as $\sim 10^{-5}$, tides may generate more heat
per unit surface area than on Io. Such an eccentricity can be
maintained by perturbations from planet c, the $8.4 \mearth$ planet
that orbits at 0.046 AU, but only if its own eccentricity is
large enough. In many plausible cases, the tidal heating may be orders of
magnitude larger.  In $\S$ 2 we discuss tidal theory, the two models
for the surface temperature presented in LRS09, and our
N-body model. In $\S$ 3 we show the possible orbital and thermal
histories of planet b. Finally, in $\S$ 4 we discuss the plausible
range of physical properties that this planet has in the context of
tidal evolution and tidal heating, and suggest that CoRoT-7 b could be the
first of many super-Ios that will be be discovered by \textit{CoRoT}
and \textit{Kepler}.

\section{Methods}
Previous investigations have considered the possible range of surface
temperatures of CoRoT-7 b (LRS09, Schaefer \& Fegley 2009; Valencia
\etal 2009). Here we review the two models of LRS09: one in which the
energy is uniformly distributed over the surface,
\begin{equation}
T_u = (1 - A)^{1/4}g\Big(\frac{R_*}{2a}\Big)^{1/2}T_*, 
\label{eq:tgr}
\end{equation}
and one in which there is no surface heat transport to the back side, 
\begin{equation}
T_s = (1-A)^{1/4}g\Big(\frac{R_*}{a}\Big)^{1/2}T_*. 
\label{eq:tss}
\end{equation}
Here, $A$ is the albedo, $g$ quantifies the effectiveness of heat
transport, and $T_*$ is the effective temperature of the star (5275 K
for CoRoT-7). We assumed $A = 0$ and $g = 1$ (no albedo and no
greenhouse effect [LRS09]). In the next section we will couple these
models, Eqs.~(\ref{eq:tgr} --\ref{eq:tss}), to the tidal evolution of
b's orbit in order to estimate its surface temperature history.

Tidal theory is a notoriously complicated and uncertain field. With
few examples of tidal evolution in the Solar System as well as the
long timescales associated with this phenomenon, firm observational
constraints are rare (see Lainey \etal [2009] for a recent
example). For more complete reviews of tidal models, consult
Ferraz-Mello \etal (2008) or Heller \etal (2009). Here we employ a
standard model that was developed to study orbital evolution of
satellites in our solar system (Goldreich \& Soter 1966; see also
Jackson \etal 2008a; Barnes \etal 2009a,b; Greenberg 2009). In this
model, a planet on a circular orbit migrates ($a$ evolves) at a rate
\begin{equation}
\label{eq:adot}
\frac{da}{dt} = -\frac{9}{2}\frac{\sqrt{G/M_*}R_*^5m_p}{Q'_*}a^{-11/2},
\end{equation}
where $G$ is the gravitational constant, $R_p$ is the planetary
radius, $R_*$ is the stellar radius, and $Q'_*$ is the star's ``tidal
dissipation function'' which encapsulates the physical response of the
body to tides, including the Love number. In order to constrain a
planet's orbital history, we integrate
Eq.~(\ref{eq:adot}) backward in time by flipping the sign. We use a
timestep of 1000 years which convergence tests demonstrated resolves the evolution. We will use this equation to
consider the past evolution of planet b's orbit.

We will also consider the possibility that the planet has a nonzero
eccentricity today in order to assess tidal heating, which is
parameterized as
\begin{equation}
\label{eq:heat}
H = \frac{63}{4}\frac{(GM_*)^{3/2}M_*R_p^5}{Q'_p}a^{-15/2}e^2,
\end{equation}
where $Q'_p$ is the planet's tidal dissipation function and $M_*$ is
the stellar mass (Peale \etal 1979; Jackson \etal 2008b). Note that if the planet has a nonzero eccentricity, Eq.~(\ref{eq:adot}) will have an additional term, but for the small values we consider below, this change is negligible.

In order to assess the surface effects of tidal heating, we can
compare our results with the processes and characteristics observed
on planetary bodies in our own solar system.  For comparison, heating
rates must be scaled to account for the differences in sizes among
bodies.  Because we are comparing effects at the surface, we consider
the heating rate per unit surface area (see \eg Williams \etal
1997; Jackson \etal 2008b,c), \ie the heat flux, $h = H/4\pi R_p^2$,
through the planetary surface. On Io, $h$ = 2 W m$^{-2}$ (from tidal heating)
(McEwen \etal 2004), resulting in intense global volcanism and a
lithosphere recycling timescale of 142 to $3.6 \times 10^5$ years (Blaney
\etal 1995; McEwen \etal 2004). On Earth, the heating comes from
the radioactive decay of U and K and produces a heat flux of 0.08 W m$^{-2}$
(Davies 1999), which is adequate for plate tectonics and some
volcanism, which is modest relative to Io.  L\'eger \etal (in prep.)
suggest the radiogenic heat flux of CoRoT-7 b is $\sim 0.3$ W m$^{-2}$.  The
actual scaling of geophysical processes among various sized planets
is much more complex than can be represented by the single parameter
$h$.  The internal effects will depend on, and modify, composition and
structure.  Qualitative differences in heat transport, such as the
roles of convection vs.~conduction, will make for complex diversity. 
Many of these differences may scale as the heating rate per unit
mass, for example. However, modeling these details would require
complicated internal geophysical simulations which are beyond the
scope of this investigation.  Here we assume that the surface heat
flux $h$ gives an adequate first-order qualitative representation of
surface effects.  Thus, we call planets ``super-Ios'' if their flux exceeds
Io's value ($h > 2$ W m$^{-2}$).  Otherwise they can be considered as
``super-Earths'' (Jackson \etal 2008c; Barnes \etal 2009a,b). We use Eq.~(\ref{eq:heat}) to make these distinctions in the following section.  

We also consider the perturbations from planet c which can maintain
b's eccentricity in a manner similar to the mutual perturbations of
the planets in the HD 40307 system (Barnes \etal 2009a). We model the
interactions with the N-body code
HNBody\footnote{http://janus.astro.umd.edu/HNBody/}, which includes
general relativity precession. We assume the orbits are coplanar, with
planet c's mass, $m_c = 8.93 \mearth$ (Queloz \etal 2009). For the
remaining orbital parameters, we used the nominal values from Queloz
\etal (2009), except the eccentricity of planet c was chosen from the
possible range $3 \times 10^{-5} < e_c < 10^{-3}$. We integrated the
orbital motion for $10^5$ years. Then, for each case,
using the average astrocentric eccentricity of b, we computed the
tidal heat flux of planet b, $h_b$. We calculate the average heating
flux because the mantle (if such a layer exists) of CoRoT-7 b probably
can not respond to heating variations with timescales of thousands of
years. We have not tested system stability numerically, but note that
the above parameter space is Hill stable (Marchal \& Bozis 1982;
Gladman 1993), which implies Lagrange stability as well (Barnes \&
Greenberg 2006).

\section{Results}
\subsection{Surface Temperature Evolution}

Here we consider the past evolution of this planet in order to
estimate how the surface temperature has changed. We consider a range
of $Q'_*$ values from $10^5$ -- $10^7$ (Mathieu 1994; Lin \etal 1996;
Ogilvie \& Lin 2002; Jackson \etal 2009a), and integrate back in time
over the past 2.5 Gyr in order to cover the full age of the system,
1.2 -- 2.3 Gyr (LRS09). Here we assume that $e_b$ has always been
small enough that tides raised on the planet have not affected the
orbital evolution (note that this assumption assumes the least amount
of migration). The evolution of the semi-major axis of planet b's
orbit is shown in the top panel of Fig.~\ref{fig:heat}. For low
$Q'_*$, the planet could have begun nearly twice as far out, whereas
if $Q'_* = 10^7$ or greater, there has been little change. The actual
value of $Q'_*$ remains uncertain.

\begin{figure}
\includegraphics[width=\textwidth]{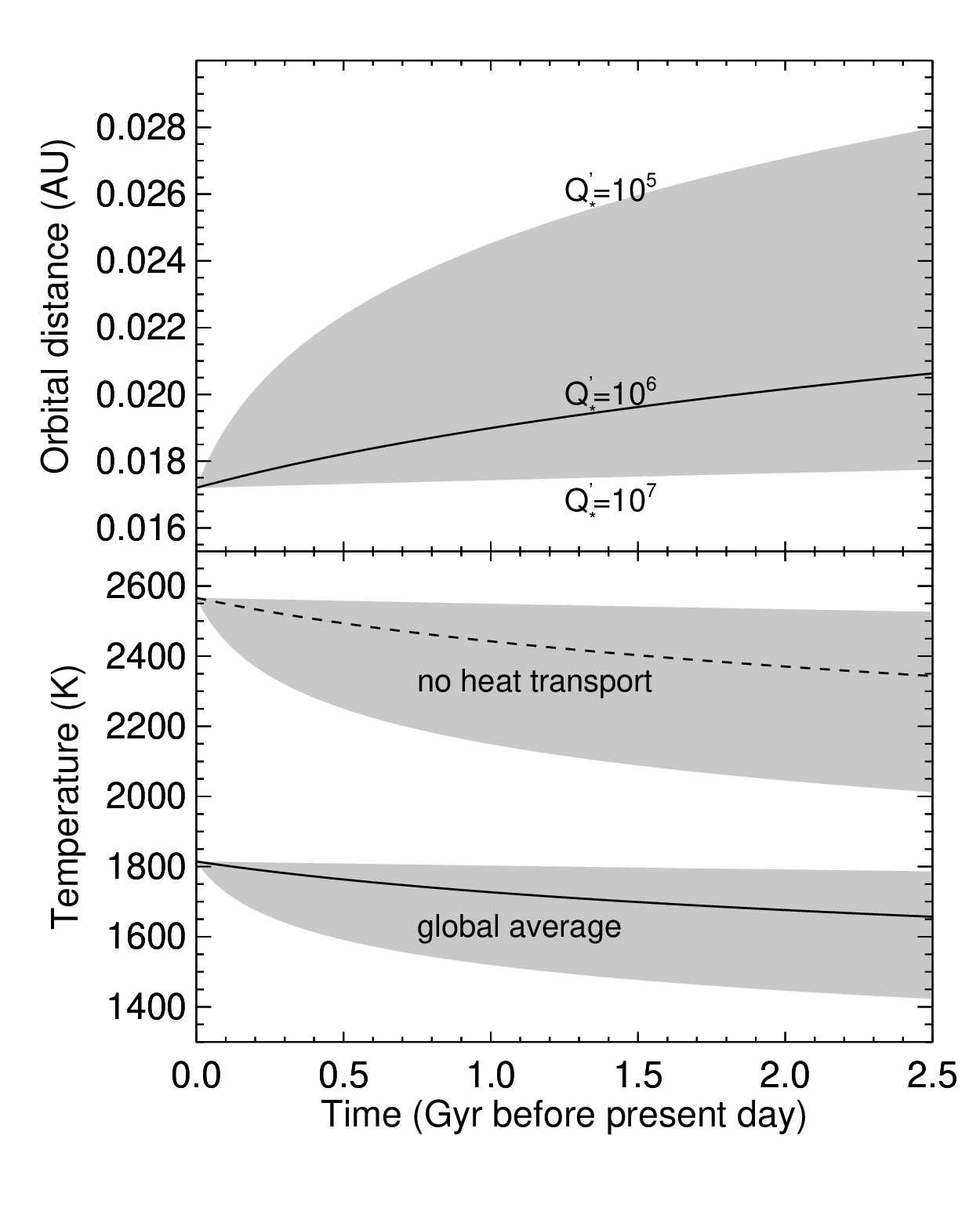}
\caption{History of CoRoT-7 b's semi-major axis (top) and surface
temperature (bottom). Note that the present time is at the left. The top panel shows a range of possible
semi-major axes due to tidal evolution for a range of $Q'_*$
values. The solid curve represents $Q'_* = 10^6$, and the gray region shows
all values in the range $10^5 \le Q'_* \le 10^7$. The bottom panel
shows the surface temperature evolution for the same range of $Q'_*$,
assuming uniform redistribution of stellar heat over the entire globe
(Eq.~(\ref{eq:tgr}); ``global average'' strip) and no
redistribution (Eq.~(\ref{eq:tss}); ``no heat transport'' strip). In
this panel, the top of the gray strips correspond to $Q'_* = 10^7$ and
the bottom to $Q'_* =10^5$.\label{fig:temp}}
\end{figure}

This evolution has important consequences for the surface temperature,
as given by Eqs.~(\ref{eq:tgr} -- \ref{eq:tss}).  Figure
\ref{fig:temp} shows the possible histories of this quantity. For
$Q'_* = 10^5$ (the lowest extent of the gray strips), the planet has
only recently reached the extreme temperature it has today (the global
average was 150 K cooler $10^8$ years ago).  Only if $Q'_*
\ge 10^7$ has the temperature been as great over the past billion years as
it is now.

\subsection{Current Tidal Heating}
\begin{figure}
\includegraphics[width=\textwidth]{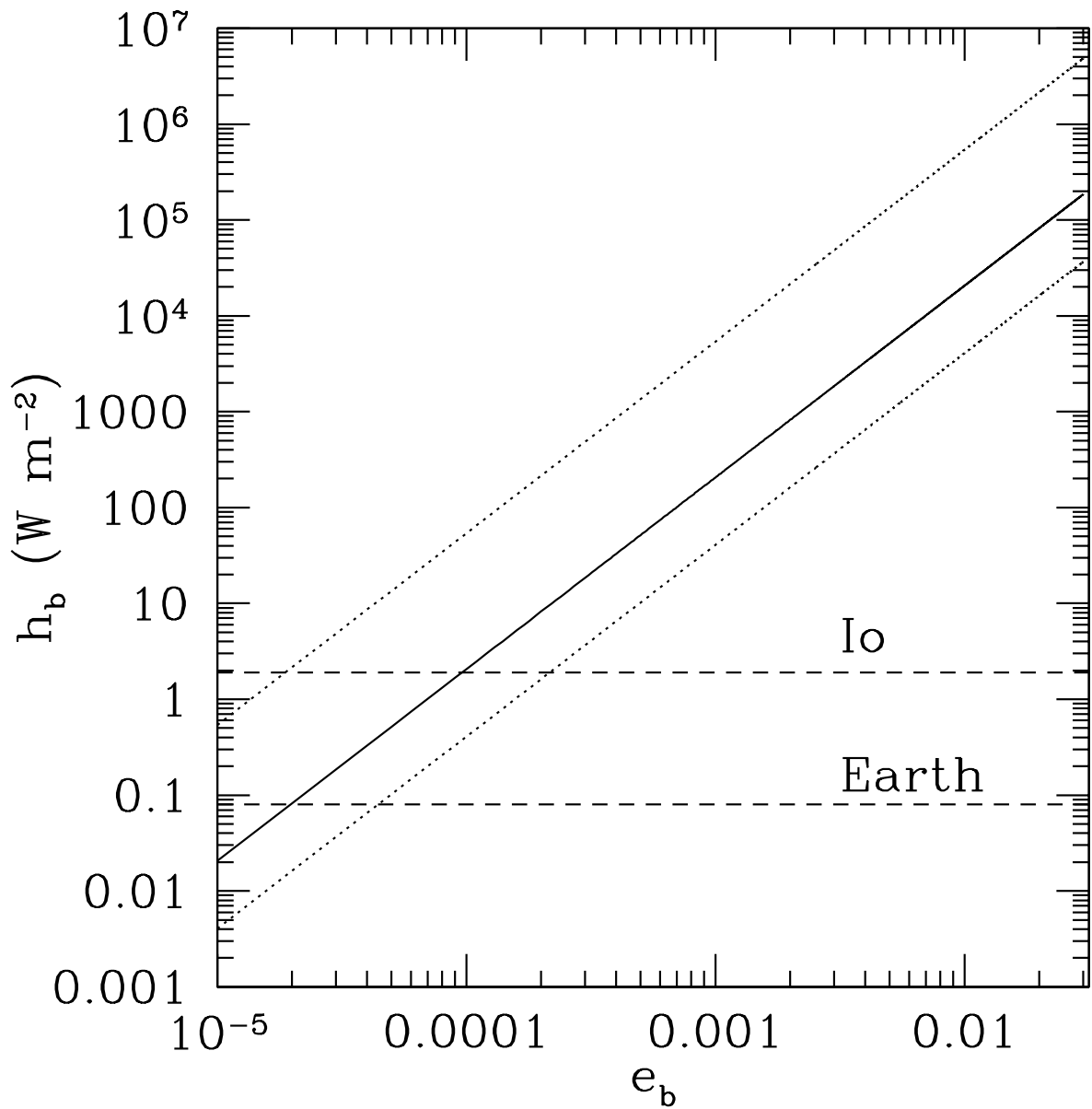}
\caption{Tidal heating of CoRoT-7 b as a function of its current eccentricity. The solid line shows the heating flux assuming nominal system parameters. The dotted lines represent a reasonable range of uncertainties: The upper line corresponds to $R_p = 1.95 \rearth$ and $Q'_p = 20$, the lower to $R_p = 1.41 \rearth$ and $Q'_p = 1000$. For reference, the tidal heating flux of Io and the radiogenic heating flux of the Earth are also shown.}
\label{fig:heat}
\end{figure}

In this subsection we explore a range of orbital configurations permitted
by the observations in order to determine which cases predict CoRoT-7
b is a super-Io and which predict a super-Earth. We begin by applying
Eq.~(\ref{eq:heat}) to find $h_b$ as a function of
$e_b$. We consider a plausible range of physical and orbital
properties listed in LRS09 (their Tables 5 and 6), with radii from 1.41 to
1.95 $\rearth$, and $Q'_p$ values from 20 -- 1000, consistent with
observations in our Solar System (Yoder 1995; Dickey \etal 1997;
Mardling \& Lin 2002; Lainey
\etal 2009). The solid line in
Fig.~\ref{fig:heat} shows the possible values of $h_b$ as a
function of its current eccentricity $e_b$, assuming the nominal system
parameters and $Q'_p = 500$. The dotted lines show a range of
possible values given the uncertainties in $Q'_p$ and $R_p$. For reference
the current heat fluxes for the Earth and Io are also shown (also note
that the stellar energy flux at the planet is $\sim 2.5 \times 10^6$ W
m$^{-2}$).

Figure \ref{fig:heat} shows that, if $e_b >
10^{-4}$ and assuming nominal radius and $Q'_p$ values, the planet is experiencing at least
as much tidal heating (in terms of surface flux) as Io.  For other plausible
parameters, the critical value of $e_b$ could be as low as $10^{-5}$.
Based on the example of our solar system, which has much larger $e$
values, one would expect that b is
indeed a super-Io.  However, without an external perturber, tides
could have damped even a very large primordial $e$ below these values in
only $10^8$ years. (Even the effects of stellar oblateness,
passing stars, and the galactic tide could not keep $e_b > 10^{-4}$ as
tidal damping occurs.)

\begin{figure}
\includegraphics[width=\textwidth]{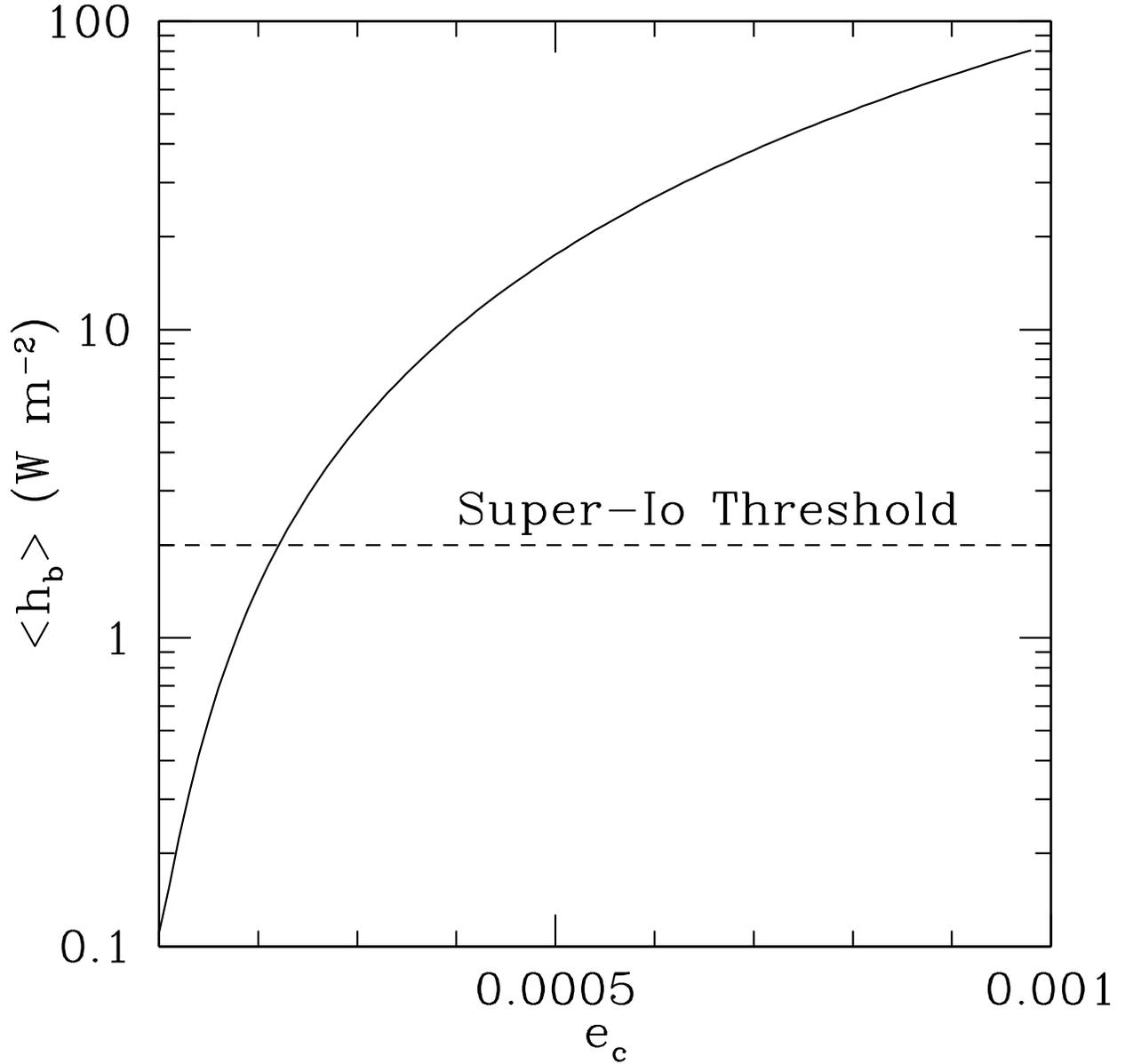}
\caption{Tidal heating of planet b resulting from 
perturbations due to planet c assuming its orbit is coplanar with b
(its mass is then $8.93 \mearth$). If $e_c \gsim 2 \times 10^{-4}$
planet b is heated at least as much as Io.}
\label{fig:driven}
\end{figure}

The most likely mechanism, therefore, for $e_b$ to be $\ge 10^{-4}$ is
through gravitational interactions with other planets in the
system. So far only one other planet is known, so we consider its
effects on planet b. The
magnitude of $e_b$ will depend on the current values of $e_c$ and
$m_c$, both of which are poorly constrained. We assume the orbit of
planet c is inclined $20^\circ$ to the line of sight (coplanar with b
[LRS09]), with a mass of $8.93 \mearth$. Fig.~\ref{fig:driven} shows
the average heating flux $<h_b>$, based on the average value of $e_b$
over the $10^5$ year integration, as a function of $e_c$. If $e_c
\gsim 10^{-4}$, then $<h_b>$ exceeds the threshold for planet b to be
a super-Io. 

But could planet c have such a ``large'' eccentricity given the extreme 
tidal damping in this system? The orbital circularization rate falls off 
as $a^{-6.5}$ (Goldreich \& Soter 1966), which might seem to suggest that 
c's orbital eccentricity would be damped relatively slowly. However, 
planets b and c are coupled through their secular interactions, so that 
the tidal damping of one may affect the other. Planet c's orbit could be 
circularized quickly through its interaction with b, i.e. planet b 
essentially drains away c's excess eccentricity. Therefore planet c alone 
may not be able to maintain b's eccentricity to the level required for it 
to be a super-Io.

\section{Discussion}
We have shown how tides may revise previous interpretations of the
physical properties of CoRoT-7 b. Consideration of tidal migration has
important implications for the planet's surface temperature, and hence
the possibility that the surface is (partially) melted (Schaefer \&
Fegley 2009). Valencia \etal (2009) argue that such a high surface
temperature would lead to loss of the volatile inventory within $10^8$
years after formation. However, Fig.~\ref{fig:temp} shows that such a
supposition implicitly assumes $Q'_* \gsim 10^6$. Although the
temperature difference could be only a few hundred degrees or less, we
encourage future research into the surface and interior properties of
CoRoT-7 b (and close-in terrestrial bodies discovered in the future)
due to insolation to bear in mind the tidal evolution of the orbit.

Previous interpretations of the surface properties of CoRoT-7 b
(L\'eger \etal 2009; Schaefer \& Fegley 2009; Valencia \etal 2009)
have only considered the radiative properties of the star. We have
shown that other stellar properties, specifically its physical
response to tides, are at least as important (at least on the dark
side of the planet). In this case the range of plausible $Q'_*$ values
permit a wide range of surface temperature histories. This result
emphasizes the urgent need to determine $Q'_*$ values more precisely.

Furthermore, we have shown that a slight eccentricity in the orbit of CoRoT-7 b can
result in enough tidal heating for it to be a super-Io planet (Barnes
\etal 2009a), even without taking into account surface melting
resulting from stellar radiation.  However, tides would quickly
circularize the orbit without an
external agent. Perturbations from CoRoT-7 c are a likely
mechanism to drive the requisite eccentricity. Depending on CoRoT-7 c's mass and orbit, CoRoT-7 b could be
tidally heated to be a super-Io or less heated so as to be a
super-Earth. Crucially, these distinctions lie below the current
detection limit for the orbital eccentricities, so either is
possible. 
However, given their close proximities to the host star, tidal damping may 
be strong enough 
to  circularize both orbits below the threshold to make b a super-Io.
Nevertheless, as shown in Fig.~\ref{fig:heat}, the uncertainties in $e_b$ are consistent with a tidal heating rate
of greater than $10^6$ W m$^{-2}$. Such a large heating rate is unlike
anything observed in our Solar System, and would have dramatic effects
on the internal, surface and atmospheric properties.

Regardless of the current eccentricity of CoRoT-7 b, most models of
planet formation predict that planets form with eccentricities of
at least a few percent (Raymond \etal 2004, 2006; O'Brien
\etal 2006; Morishima \etal 2008). Therefore it seems likely that the
planet was a super-Io early in its history (assuming it never had a
large gaseous envelope), even accounting for tidal
migration which implies formation was somewhat farther from the star
than the current orbit.

We have assumed that CoRoT-7 b has been a terrestrial-like planet
throughout its history. Valencia \etal (2009) point out that ablation
could strip away many Earth masses of gas within 1 Gyr, but we have
showed it used to be further out. We are currently examining mass loss
with tidal evolution (Jackson \etal, in prep.). If the planet began
with a large gaseous envelope, many of the ideas presented here may
require revision.

Tides play a crucial role in defining the characteristics of the atmosphere, surface and interior of CoRoT-7 b. This planet is
the first close-in terrestrial planet discovered, but the \textit{CoRoT}
and \textit{Kepler} spacecraft are likely to find many more. Mayor
\etal (2009a) find that 30\% of solar-mass stars have super-Earth to
Neptune mass companions on periods of 50 days or less, Marcy \etal
(2005) and Cumming \etal (2008) find that the mass-frequency
distribution of exoplanets is a power law with slope -1.1 (low mass
planets are more common), and the geometric transit probability of a
10 $\mearth$ planet orbiting a 1 $\msun$ star is $\sim 10$\%. Taken
together these results suggest \textit{Kepler}, which is monitoring
$10^5$ stars, should find at least 100 terrestrial planets with
orbital periods less than 10 days. Like CoRoT-7 b, these planets can
be strongly tidally heated. A significant fraction of the first wave
of rocky exoplanets may be super-Ios.

\section{Acknowledgments}
RB and SNR acknowledge funding from NASA Astrobiology Institute's
Virtual Planetary Laboratory lead team, supported by NASA under
Cooperative Agreement No. NNH05ZDA001C. RG acknowledges support from
NASA's Planetary Geology and Geophysics program, grant
No. NNG05GH65G. BJ is funded by an NPP administered by ORNL.

\clearpage

\clearpage

\end{document}